\documentclass[twocolumn,showpacs,preprintnumbers,amsmath,amssymb,prl,superscriptaddress]{revtex4}

\begin{document}

\title{Comment on ``Projective Quantum Monte Carlo Method for the
Anderson Impurity Model and its Application to Dynamical Mean
Field Theory''}

\author{M. I. Katsnelson}
\email{m.katsnelson@science.ru.nl} \affiliation{Institute for
Molecules and Materials, Radboud University Nijmegen, 6525 ED
Nijmegen, The Netherlands} 

\begin{abstract}
\end{abstract}

\maketitle

In a recent Letter \cite{held}, Feldbacher, Held, and Assaad have
proposed a new scheme for the Quantum Monte Carlo simulations of
the impurity Anderson model. This method is supposed to calculate
the average of an operator $\mathcal{O}$ over the ground state of
the Hamiltonian $H$ via the expression
\begin{equation}
\left\langle \mathcal{O}\right\rangle _{0}=\lim_{\theta
\rightarrow \infty }\lim_{\beta \rightarrow \infty }\frac{Tr\left[
e^{-\beta H_{T}}e^{-\theta H/2}\mathcal{O}e^{-\theta H/2}\right]
}{Tr\left[ e^{-\beta H_{T}}e^{-\theta H}\right] }  \label{aver}
\end{equation}
where $H_{T}$ is an appropriate trial Hamiltonian (see
Ref.\onlinecite{held}, Eq.(3)). The main requirement to the choice
of $H_{T}$ is that its ground state can not be orthogonal to the
ground state of the Hamiltonian under consideration $H$. In real
calculations, the authors of Ref.\onlinecite{held} have taken $H$
as the impurity Anderson Hamiltonian and $H_{T}$ as its
noninteracting part, i.e. the Anderson Hamiltonian without the
Hubbard interaction term $-U/2\left ( n_{\uparrow}-n_{\downarrow}
\right )^2$ where $n_{\sigma }$ is the operator of number of
localized electrons with spin projection $\sigma
=\uparrow,\downarrow$ on the impurity site. Unfortunately, this
choice is not consistent with the basic restriction mentioned
above due to the Anderson orthogonality catastrophe
\cite{anderson,VK}. Anderson has proven that in thermodynamic
limit $N\rightarrow \infty$ (where $N$ is the number of electrons)
a local perturbation leads to complete reconstruction of the
ground state of fermionic system in such a way that the overlap of
the ``old'' and ``new'' ground-state way functions, $\left| \Psi
_{T}\right\rangle $ \ and $\left| \Psi \right\rangle $, is
proportional to $N^{-\alpha }$ where for the case of spherically
symmetric perturbation potential
\[
\alpha =\sum\limits_{l=0}^{\infty }(2l+1)\left( \frac{\delta _{l}}{\pi }%
\right) ^{2},
\]
$\delta _{l}$ being the scattering phases at the Fermi level with
orbital quantum number $l$. Later this expression was generalized
to the case of arbitrary local perturbation \cite{YY79}:
\[
\alpha =\frac{1}{2}\left( \frac{1}{2\pi i}\right) ^{2}Tr\left( \ln \widehat{S%
}\right) ^{2}
\]
where $\widehat{S}$ is the scattering matrix of the local
perturbation at the Fermi surface. Original derivation of the
Anderson orthogonality catastrophe was done for the case of local
one-body potential; however, later it has been proven also for the
Hubbard interaction term \cite{YY78}. Taking into account the
Friedel sum rule for the Anderson model \cite{Langreth,Hewson} one
can demonstrate that $\alpha$ is proportional to $\left (
n_{d}-n^{0}_{d} \right )^2$ where $n_{d}$, $n^{0}_{d}$ are the
impurity occupation numbers for the Hamiltonians $H$ and $H_T$,
correspondingly. For a special case of symmetric half-filled
Anderson model one has $\alpha = 0$ \cite{reply}; however, to
satisfy this in a generic case one has to know already an exact
answer for $n_{d}$ which makes the whole scheme unpractical, at
least, in the context of the dynamical mean-field theory. For the
case of non-degenerate Anderson model the value $n_{d}$ can be
calculated exactly by the Bethe Ansatz method \cite{Wiegmann}.

Thus, for the infinite system the ground state does not contribute
to both numerator and denominator of the ratio (\ref{aver}) and,
instead of being the average over the ground state the latter is
rather some weighted average over excited states.

The ``orthogonality catastrophe'' is a generic and fundamental
property of many-body fermionic systems (see, e.g., discussion in
Ref.\onlinecite{book}) and should be carefully taken into account
at the development of different projection techniques.

\end{document}